\documentclass[12pt]{article}

\usepackage[totalheight = 23cm, totalwidth = 17cm]{geometry}
\usepackage{amssymb,amsmath,amsfonts,amsbsy,graphicx}
\usepackage{color}  
\usepackage[symbol]{footmisc}
 \usepackage{ulem}
 
\newcommand{\mpl}{m_{\rm Pl}}
\newcommand{\calO}{{\cal O}}
\newcommand{\calR}{{\cal R}}

\begin{document}

\begin{titlepage}

\rightline{\footnotesize{APCTP-Pre2021-001}} 
\vspace{-0.3em}
\rightline{\footnotesize{PNUTP-21-A11}}

\begin{center}

\vskip 3em

{\LARGE \bf 
Axion-driven hybrid inflation over a barrier
}

\vskip 3em

{\large
Jinn-Ouk Gong$^{a,b}$ 
and
Kwang Sik Jeong$^{c}$ 
}

\vskip 0.5cm

{\it
$^{a}$Department of Science Education,  Ewha Womans University, Seoul 03760,  Korea
\\
$^{b}$Asia Pacific Center for Theoretical Physics,  Pohang 37673,  Korea
\\
$^{c}$Department of Physics,  Pusan National University,  Busan 46241,  Korea
}

\vskip 1.2cm

\end{center}

\begin{abstract}

We present a novel cosmological scenario that describes both inflation and dark matter.
A concrete realization of our scenario is given based on a well-established particle physics model, 
where an axionlike field drives inflation until a potential barrier, 
which keeps a waterfall field at the origin, disappears to trigger a waterfall transition. 
Such a barrier makes the inflaton potential much flatter, improving significantly the naturalness 
and viability of the otherwise problematic setup adopted previously. 
The observed spectrum of the cosmic microwave background indicates that 
the inflationary Hubble scale, which is allowed to span a wide range, uniquely fixes the inflaton mass 
and decay constant. 
This raises an intriguing possibility of probing inflation via experimental searches for axionlike particles. 
Further, our model involves dark matter candidates including the inflaton itself. 
Also, for a complex waterfall field, we can determine cosmologically 
the Peccei-Quinn scale associated with the strong CP problem.

\end{abstract}

\end{titlepage}

\newpage

\section{INTRODUCTION}

Cosmic inflation~\cite{Guth:1980zm,Linde:1981mu,Albrecht:1982wi}
has become an essential part of the standard cosmological model. 
Before the onset of the hot big bang evolution, 
it provides the necessary initial conditions--otherwise extremely finely tuned--as confirmed 
by the observations on the cosmic microwave background (CMB)~\cite{Aghanim:2018eyx}.
Furthermore,  
it explains the origin of temperature fluctuations of the CMB and the inhomogeneous distribution 
of galaxies on large scales due to quantum fluctuations during inflation~\cite{Mukhanov:1981xt}. 
The properties of these primordial perturbations have been constrained by decades of observations, 
and are consistent with the predictions of inflation~\cite{Akrami:2018odb}.

To implement inflation, we need typically an inflaton field with a sufficiently flat potential.
The inflaton drives an inflationary epoch until the ``slow-roll'' period does not hold any 
longer~\cite{Mukhanov:2005sc,Weinberg:2008zzc}. 
It is, however, a formidable task to maintain an unusually flat potential against various 
corrections~\cite{Lyth:1998xn}. 
A powerful way of protecting the flatness is to impose certain symmetries. 
An axionlike field is an appealing candidate for an inflaton because its mass requires breaking 
of the associated shift symmetry,  naturally making it very light.  
However,   the predictions of the minimal axion-driven inflation -- natural 
inflation~\cite{Freese:1990rb,Adams:1992bn} -- are only marginally consistent with the most 
recent Planck observations. 
Additionally,  for successful inflation,  the axion should 
have a trans-Planckian decay constant $f \gtrsim 5\mpl$ with 
$\mpl \equiv (8\pi G)^{-1/2}$~\cite{Akrami:2018odb}, 
which may be outside the range of validity of an effective field theory description.

In this article, we present a novel cosmological scenario in which an axionlike field  
can drive inflation successfully and at the same time contribute to dark matter. 
The end of the inflationary phase is triggered 
by a waterfall transition like hybrid inflation~\cite{Linde:1993cn}. 
The distinctive features of our model are twofold. 
First,  the waterfall field $\chi$ is trapped at the origin during inflation by a potential barrier. 
This implies that, differently from the previous hybrid inflation models,   
the scale of inflation is not tied to that of the waterfall phase transition. 
As a result, unlike the original natural inflation, the inflaton is allowed to have a decay constant 
well below $\mpl$ so that the effective field theory is trustable,   yet maintains a flat potential. 
Thus the naturalness and viability of the models, which were problematic either 
in the effective theory viewpoint or in unnaturally fine-tuned initial conditions, 
are improved significantly within our scenario.
The Planck results are accommodated in a broad range of the inflaton mass and decay constant, 
but with a certain relationship between them. 
This opens an interesting possibility to probe inflation via experimental searches for axionlike particles.

Another merit of our scenario is that the inflaton itself can constitute dark matter,
which is generally difficult in other inflation models.  
Further, because the inflationary Hubble scale $H_{\rm inf}$ is allowed to span a very wide range,
$\chi$ has a potential to resolve other puzzles of the Standard Model (SM). 
If complex, we may identify U$(1)_\chi$ with the Peccei-Quinn (PQ) symmetry
solving the strong CP problem~\cite{Peccei:1977hh}. 
Remarkably, the PQ scale is then determined cosmologically,   
and the contribution of the QCD axion to dark matter constrains 
$H_{\rm inf}$ to be below about $10^4$~GeV.

\section{MODEL}
\label{sec:model}

Our model consists of two scalar fields,  the inflaton $\phi$ and the waterfall field $\chi$. 
During inflation, $\phi$ rolls down the potential slowly while $\chi$ is trapped 
at the origin by a barrier.
There, the effective mass squared of $\chi$, $\mu_{\rm eff}^2$, is thus positive.
As $\phi$ evolves, $\mu_{\rm eff}^2$ 
decreases monotonically and vanishes at a critical value $\phi_c$, removing the barrier.
Then inflation ends almost instantaneously.
Here, the barrier does bring the separation between the scales for inflation and waterfall phase transition. 

Our scenario is successfully realized if $\phi$ is an axionlike field with decay constant $f$.
This is because its interactions are well controlled by 
shift symmetry,  $\phi\to \phi+ {\rm constant}$,
presumably broken only by nonperturbative effects, 
and the size of its potential terms is finite and insensitive to $f$.  
A dangerous waterfall tadpole can be avoided by imposing a symmetry,  for instance, 
U$(1)_\chi$ if $\chi$ is a complex scalar. 
Explicitly,   we consider the potential
\begin{equation}
\label{eq:full-potential}
V(\phi,\chi) = V_0 + \mu^2_{\rm eff}(\phi) |\chi|^2 - \lambda |\chi|^4 + \frac{1}{\Lambda^2} |\chi|^6
+ U(\phi) \, ,
\end{equation} 
with $\lambda>0$, and the $\phi$-dependent terms 
given respectively by
\begin{align}
\label{eq:mass-squared}
\mu^2_{\rm eff}(\phi) 
& = 
m^2 - \mu^2 \cos\left( \frac{\phi}{f} + \alpha \right)
\, ,
\\
\label{eq:inflaton-potential}
U(\phi) 
& =
M^4 \cos\left( \frac{\phi}{f} \right)
\, ,
\end{align}
where $\alpha$ is constant.
The positive constant $V_0$ is fixed by demanding $V=0$ at the true vacuum,  
and $\Lambda$ is the cutoff scale of the theory. 
The parameter space of our interest is
\begin{equation}
\label{eq:parameter-space}  
m^4 < \mu^4 \ll \lambda V_0
\quad \text{and} \quad
M^4 \ll V_0 \, .
\end{equation}
%
%
Then the true vacuum appears at $\chi_0 \sim \sqrt{\lambda} \Lambda$, 
well below the cutoff scale $\Lambda$ as long as $\lambda\ll 1$, 
and $V_0$ reads
\begin{equation}
V_0 \sim \lambda^3 \Lambda^4 \, . 
\end{equation}
Note that there are two minima along the $\chi$-direction for $\mu^2_{\rm eff}(\phi)>0$, 
and a barrier separates them.
The position and height of the waterfall barrier are determined by $\mu$,  whereas
the value of $V_0$ is insensitive to it.

Before going further, let us discuss the case $\lambda<0$ so that there is no barrier,
similar to hybrid natural inflation~\cite{Ross:2009hg,Ross:2010fg,Ross:2016hyb}.
In such a case,  $V_0$ is fixed by $\mu$ roughly to be $\mu^4/|\lambda|$,   and
the possible range of $M^4/V_0$ is severely constrained because a closed loop of $\chi$
generally makes 
$|\lambda|\gtrsim 1/16\pi^2$ and $M^4 \gtrsim \mu^2 \Lambda^2_\ast /16\pi^2$
with $\mu<\Lambda_\ast$.
Here $\Lambda_\ast$ is the cutoff scale of the $\phi$-dependent waterfall mass
operator.   
In our scenario, $M^4/V_0$ can be arbitrarily small,  making the inflaton potential
much flatter than the case without a barrier. 

The rate of tunneling over a barrier is proportional to $\exp(-S_E)$,  
where $S_E$ is the Euclidean action of $\chi$ evaluated on a bounce solution.  
Tunneling proceeds dominantly via the Coleman-De Luccia bounce~\cite{Coleman:1980aw} with 
$S_E>S_0\equiv 8\pi^2/(3\lambda)$ in the region with $\mu^2_{\rm eff}> 2H^2_{\rm inf}$, 
while through the Hawking-Moss instantons~\cite{Hawking:1981fz} with 
$S_E= \mu^4_{\rm eff}/H^4_{\rm inf} \times S_0$ in the opposite region~\cite{Shkerin:2015exa}.  
Here we have used that the bounce is insensitive to the $|\chi|^6$ term for 
$\mu^4 \ll \lambda V_0$. 
For viable inflation,  we thus impose the condition
\begin{equation} 
\label{eq:tunneling-suppression} 
\mu^2 \gg H_{\rm inf}^2 \, .
\end{equation}
Then $\chi$ is heavy enough to be initially fixed at the origin.  
In addition, the tunneling rate is exponentially suppressed 
so that $\chi$ stays at the origin until the barrier disappears at $\phi_c$.
Bubbles of true vacuum can be nucleated around the end of inflation,
but the U$(1)_\chi$ phase transition occurs rather
smoothly because the barrier soon disappears.

As a simple ultraviolet completion of the inflaton potential, we consider 
a hidden QCD with U$(1)_\chi$ charged quarks.
\eqref{eq:mass-squared} and \eqref{eq:inflaton-potential} are then
generated in a controllable way
while naturally satisfying the hierarchies \eqref{eq:parameter-space}. 
Vectorlike quarks $u+u^c$ and $d+d^c$ couple to $\chi$ through the U$(1)_\chi$ and 
gauge invariant interactions
\begin{equation}
\label{eq:uv-completion}
m_u uu^c + y \chi u^c d + y^\prime \chi^* u d^c + m_d dd^c 
+ \frac{1}{16\pi^2}\frac{\phi}{f}\, G_{\mu\nu} \widetilde{G}^{\mu\nu}
\, ,
\end{equation}
where the hidden confining scale lies in the range $m_d \ll \Lambda_h \ll m_u$. 
Here we have taken the field basis where the quark mass parameters are real. 
Note that the last term above is an anomalous inflaton coupling to hidden gluons, 
which is the only source of shift symmetry breaking.  
At energy scales below $m_u$,   $u+u^c$ are integrated out to give
a $\chi$-dependent effective quark mass 
\begin{equation} 
\left(
\frac{yy^\prime}{m_u} |\chi|^2 + m_d + \delta m_d
\right) dd^c \, .
\end{equation}
Here we have included the radiative contribution from a closed loop of $\chi$
\begin{equation}
\delta m_d = 
\frac{yy^\prime}{16\pi^2} m_u \log \left( \frac{\Lambda^2}{m^2_\chi} \right) 
\, ,
\end{equation}
with $m_\chi$ being the mass of the radial component of $\chi$.
For small values of $\chi$,  $d+d^c$ are lighter than $\Lambda_h$
and condensate to form a meson with mass and decay constant around $\Lambda_h$.
The inflaton mixes with the meson in the presence of
anomalous coupling to hidden gluons,
and finally the effective potential at scales below $\Lambda_h$ is 
obtained by integrating out the heavy meson
\begin{equation}
\Delta V_{\rm eff} 
=
-\left| \frac{yy^\prime}{m_u}\right| \Lambda^3_h  
\cos\left( \frac{\phi}{f} + \beta_1 \right) |\chi|^2
+ 
\left| m_d  + \delta m_d \right|
\Lambda^3_h \cos\left( \frac{\phi}{f} + \beta_2  \right)
\, ,
\end{equation}
where the constant phases are given by $\beta_1 = \arg(yy^\prime/m_u)$ and 
$\beta_2 = \arg(m_d +\delta m_d)$. 
It is clear that the above reduces to \eqref{eq:mass-squared} and \eqref{eq:inflaton-potential} 
with $\alpha = \beta_1-\beta_2$. 
Also,  the hierarchies
$\mu^4 \ll \lambda V_0$ 
and $M^4\ll V_0$ are satisfied naturally if 
%
\begin{equation}
H_{\rm inf} \lesssim \Lambda_h \ll \Lambda \, ,
\end{equation}
where we have used that $H_{\rm inf}$ should be lower than $\Lambda_h$ 
since otherwise instanton effects become very weak. 
On the other hand, 
$m$ should be smaller than $\mu$ because inflation ends when the barrier disappears.
The smallness of $m$ may be accommodated in more speculative idea like supersymmetry
or anthropic selection.

\section{COSMOLOGICAL DYNAMICS}
\label{sec:cosmo}

\subsection{Inflation}

\begin{figure}[t] \centering
\includegraphics[height=0.27\textheight]{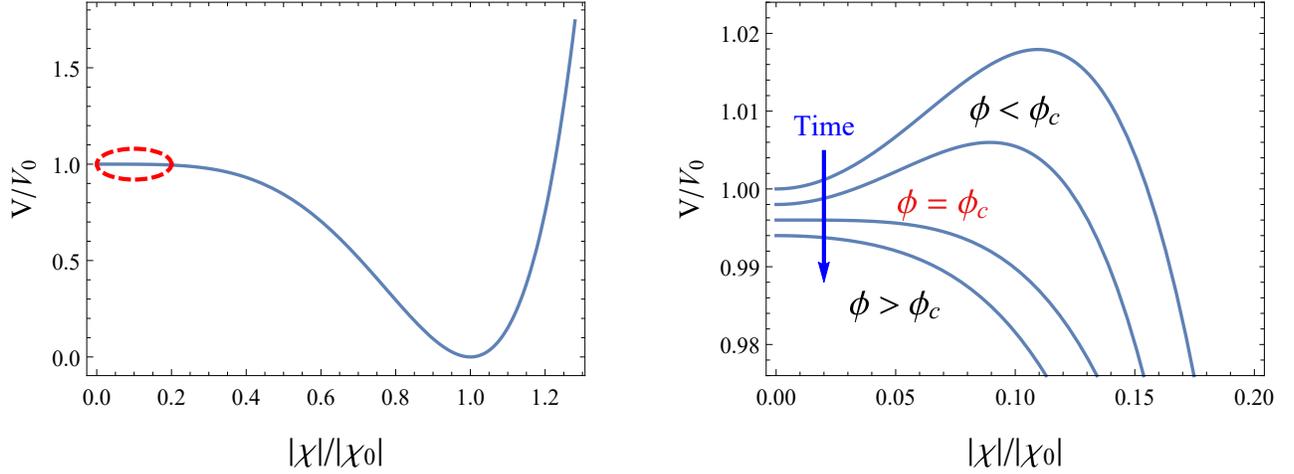}
\caption{ 
Schematic display of inflationary and waterfall phases.  
The evolution of $\phi$ changes
the waterfall potential  dramatically near the origin,  as illustrated in the right panel,
but rarely around the true vacuum at $\chi=\chi_0$ if 
\eqref{eq:parameter-space} holds.  
}
\label{fig:inflation}
\end{figure}

The Universe undergoes an inflationary phase while $\chi$ is trapped at the origin. 
During this stage, $\phi$ evolves down the potential
\begin{equation}
\label{eq:during-inf-potential}
V = V_0 + U(\phi) = V_0 + M^4 \cos\left( \frac{\phi}{f} \right) \, .
\end{equation}
Thus,  the evolution of $\phi$ during inflation is essentially identical to hybrid natural inflation. 
$\mu^2_{\rm eff}$ crosses zero when $\phi$ reaches the critical value
\begin{equation}
\label{eq:phi-critical}
\frac{\phi_c}{f} = \cos^{-1} \left( \frac{m^2}{\mu^2} \right) - \alpha \, .
\end{equation} 
The sign flip triggers the waterfall phase transition, because there is no potential barrier 
along the $\chi$-direction, 
and inflation ends almost instantaneously. 
Among the model parameters,  $m$,  $\mu$,  and $\alpha$ affect inflation only through the above combination. 
Figure~\ref{fig:inflation} shows schematically the inflationary and waterfall phases.

At this point, it is very important to note that two crucial ingredients are required 
to make our scenario distinctive.
One is the shift symmetry of $\phi$, 
which naturally allows the hierarchies \eqref{eq:parameter-space}.
The other is a $\phi$-dependent
barrier between two extrema at and off the origin in the waterfall potential. 
The barrier makes the inflaton potential flatter, and consequently 
the value of $f$ required for viable inflation can naturally be much lower than $\mpl$.

For $M^4 \ll V_0$,  inflation is driven by $V_0$ so that 
\begin{equation}
\label{eq:Hinf}
H_{\rm inf}^2 \approx \frac{V_0}{3\mpl^2} \, .
\end{equation}
From \eqref{eq:during-inf-potential}, the slow-roll parameters are given by,  
with $\theta \equiv \phi/f$,
\begin{align}
\label{eq:epsilon}
\epsilon 
& \equiv 
\frac{\mpl^2}{2} \left( \frac{V^\prime}{V} \right)^2
\approx
\frac{1}{2}
\left( \frac{\mpl}{f} \right)^2
\left( \frac{M^4}{V_0} \right)^2
\sin^2\theta
\, ,
\\
\label{eq:eta}
\eta 
& \equiv 
\mpl^2 \frac{V^{\prime\prime}}{V} 
\approx
- \left( \frac{\mpl}{f} \right)^2
\left( \frac{M^4}{V_0} \right)
\cos\theta
\, .
\end{align}
Thus,  $|\eta|$ is parametrically much bigger than $\epsilon$.
The slow-roll conditions, $\epsilon\ll 1$ and $|\eta|\ll 1$, are satisfied if the following condition holds
\begin{equation}
f \gtrsim \left( \frac{M^4}{V_0} \right)^{1/2} \mpl \, ,
\end{equation}
but it need not be above $\mpl$.

The amplitude of the power spectrum of the curvature perturbation and its spectral index, 
and the tensor-to-scalar ratio in terms of the slow-roll parameters are 
constrained as~\cite{Akrami:2018odb}
\begin{align}
\label{eq:amplitude}
A_\calR
& =
\frac{V_0}{24\pi^2\mpl^4\epsilon_*}
\approx 2.0989^{+0.0296}_{-0.0292} \times 10^{-9} 
\, ,
\\
\label{eq:index}
n_\calR 
& =
1 - 6\epsilon_* + 2\eta_*
\approx 0.9656 \pm 0.0042 
\, ,
\\
\label{eq:tensor-to-scalar}
r 
& =
16\epsilon_*
< 0.056
\, ,
\end{align}
where the subscript $*$ denotes the evaluation at the horizon exit. 
Since $\epsilon \ll |\eta|$,  $n_\calR$ is determined entirely by $\eta$. 
Hence, from \eqref{eq:eta} and \eqref{eq:index}, $f$ is written as
\begin{equation}
\label{eq:f-from-r}
f 
= 
\sqrt{\frac{2}{1-n_\calR} \cos\theta_*} \bigg( \frac{M^4}{V_0} \bigg)^{1/2} \mpl
\approx
7.625 \sqrt{\cos\theta_*} \bigg( \frac{M^4}{V_0} \bigg)^{1/2} \mpl
\, ,
\end{equation}
while \eqref{eq:tensor-to-scalar} is translated to the following mild constraint
\begin{equation}
\frac{M^4}{V_0} 
<
\frac{0.056}{8} \frac{2}{1-n_\calR} \cos\theta_*
\approx
0.4070 \cos\theta_*
\, .
\end{equation}
The number of $e$-folds $N$,  before the onset of the waterfall phase transition,   is estimated by
\begin{equation}
N 
= 
\frac{1}{\mpl} \int_{\phi_c}^\phi \frac{d\phi'}{\sqrt{2\epsilon}}
\approx
\frac{V_0}{M^4} \bigg( \frac{f}{\mpl} \bigg)^2 \log \bigg[ \frac{\tan(\theta_c/2)}{\tan(\theta/2)} \bigg]
\approx
58.14 \cos\theta_* \log \bigg[ \frac{\tan(\theta_c/2)}{\tan(\theta/2)} \bigg]
\, ,
\end{equation}
thus we can use $\theta$ and $N$ interchangeably. 
The required number of $e$-folds is around 60, 
which fixes $\theta_\ast$ roughly as
\begin{equation}
\theta_\ast \approx 
0.7126 \tan \bigg( \frac{\theta_c}{2} \bigg) - 0.1566 \tan^3 \bigg( \frac{\theta_c}{2} \bigg)\,,
\end{equation}
neglecting terms of higher order in $\tan(\theta_c/2)$.
Thus, $\theta_\ast$ does not need to be very close to the hilltop of the potential.

\begin{figure}[t] \centering
\includegraphics[height=0.27\textheight]{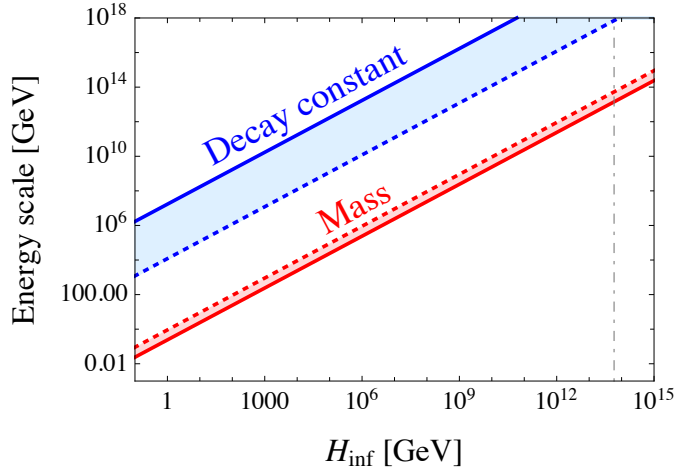}
\caption{ 
Decay constant and lower bound on mass of the inflaton
compatible with the Planck results on
$n_{\cal R}$ and $A_{\cal R}$.
We have taken $\theta_\ast$ lying between 0.01 (solid lines) and 1.5 (dotted lines).  
}
\label{fig:inflaton}
\end{figure}

The inflaton mass during inflation sets the lower bound of the inflaton
mass at the true vacuum,   $m_\phi|_{\rm min} = M^2/f$. 
It is interesting to note that both $f$ and $m_\phi|_{\rm min}$ are proportional to $H_\text{inf}$;
combining \eqref{eq:amplitude} with \eqref{eq:Hinf} and \eqref{eq:f-from-r}, 
they are written respectively as
\begin{align}
\label{eq:Hinf-from-Ar}
f
& = 
\frac{H_\text{inf}}{\pi(1-n_\calR)\sqrt{A_\calR} \tan\theta_*}
\approx
\frac{2.020 \times 10^5}{\tan\theta_*} H_\text{inf}
\, ,
\\
\label{eq:inflaton-mass-coupling}
m_\phi|_{\rm min}
& =
\sqrt{\frac{3(1-n_\calR)}{2\cos\theta_*}} H_\text{inf}
\approx
\frac{0.2272}{\sqrt{\cos\theta_*}} H_\text{inf}
\, .
\end{align}
Figure~\ref{fig:inflaton} shows the relationship between $H_{\rm inf}$,  $m_\phi|_{\rm min}$ and $f$.
$\phi$ can couple to the SM sector,   for instance,   to gauge bosons through anomalous 
couplings as naturally expected from its axionic nature.  
Our scenario thus provides theoretical support for experimental searches 
for axionlike particles in a wide mass range.
The rough relation, $f\sim 10^6 \times m_\phi|_{\rm min}$, indicates that $\phi$ should be heavier 
than about $0.1$~GeV to avoid too rapid cooling of stars~\cite{Raffelt:1999tx},  
if coupled to photons.   
Another plausible possibility is that it instead couples to the Higgs sector as 
in the cosmological relaxation model of the weak scale~\cite{Graham:2015cka}.
Then,   it may be detectable at collider and beam-dump experiments.
For instance,   the mass range below a few GeV can be probed by experiments
at SHiP~\cite{Alekhin:2015byh} and NA62~\cite{Martellotti:2015kna}.

The postinflationary evolution leads to very rich phenomenologies. 
The quantitative predictions depend very much on the detail of the model, 
so here we are satisfied with describing briefly the subsequent evolution. 
After the barrier disappears, $\chi$ soon acquires a tachyonic mass 
much larger than $H_\text{inf}$ in magnitude for $\mu^2\gg H^2_\text{inf}$.  
This happens within an $e$-fold after $\phi=\phi_c$, 
so $\chi$ rolls fast down to the true vacuum. 
Unlike usual axion-driven inflation where the Universe is reheated via an anomalous coupling to photons, 
the reheating process depends greatly on the details of the model. 
Generally speaking, however, depending on the couplings, tachyonic preheating~\cite{Felder:2000hj} is 
extremely effective so that the vacuum energy is rapidly transferred to the energy of inhomogeneous 
oscillations of $\phi$ and/or $\chi$~\cite{GarciaBellido:1997wm,Copeland:2002ku}, 
subsequently heating up the Universe to a radiation-dominated regime.

After inflation, spontaneous U$(1)_\chi$ breaking occurs and
leads to the formation of cosmic strings~\cite{Jeannerot:2003qv}, which can survive 
in the late Universe and contribute to the CMB temperature anisotropies,  
depending on how the associated Nambu-Goldstone boson becomes massive.
For instance, for global U$(1)_\chi$,  it can obtain a mass nonperturbatively from
some confining gauge sector. 
Then, topological defects are unstable and collapse by the wall tension
if the domain-wall number is equal to unity,
or if a small explicit symmetry-breaking term is added to lift the vacuum 
degeneracy~\cite{Gelmini:1988sf,Larsson:1996sp}.
Further, cosmic string loops and large time-dependent inhomogeneities generated 
during tachyonic preheating can act as a source of gravitational waves (GWs). 
The corresponding GW spectrum can span a huge range of frequencies; 
from $\calO(10^{-12})$ Hz to $\calO(1)$ Hz for stable and metastable cosmic strings~\cite{Auclair:2019wcv}, 
within the reach of pulsar-timing arrays,  LIGO, and LISA,  
and from $\calO(1)$ Hz to $\calO(10^{10})$ Hz for inhomogeneities from tachyonic 
preheating~\cite{GarciaBellido:2007dg,GarciaBellido:2007af,Dufaux:2008dn}. 
Such high-frequency GWs are unfortunately beyond the sensitivity of 
interferometric experiments due to the shot-noise fluctuations of photons. 
GWs in relatively low-frequency regimes may well be within the reach of 
future detectors like advanced LIGO,  the Einstein Telescope,  and the Big Bang Observer, 
which however is possible only for extremely small values of couplings.

\subsection{Dark matter}

Another distinctive feature of our scenario is the possibility that the inflaton can contribute 
to dark matter if its potential arises from hidden QCD as in \eqref{eq:uv-completion}.
Having Yukawa interactions with $\chi$, the hidden quarks have masses increasing with 
the waterfall field value.   
This implies that $\Lambda_h$ also increases, and thus the hidden QCD gets stronger after inflation.
In the region of large waterfall field values  where all the hidden quarks are heavier than 
$\Lambda_h$, we have
\begin{equation}
\label{eq:DM}
\mu^2= 0 
\quad \text{and} \quad
M^4 = \Lambda^4_h\,,
\end{equation}
in \eqref{eq:mass-squared} and \eqref{eq:inflaton-potential}, 
because there are no mesons formed.

Let us consider the case that $\chi_0$ is sufficiently large so that 
\eqref{eq:DM} holds in the present Universe.
$\phi$ is then stabilized at a CP-conserving minimum,  
and consequently accidental $Z_2$ symmetry arises: $\phi\to -\phi$. 
The $Z_2$ forbids $\phi$ to mix with $\chi$, 
making it stable if does not couple to SM.
$\phi$ starts coherent oscillations around the minimum when 
$H$ becomes comparable to its mass, i.e.~at the temperature fixed by
\begin{equation}
m_\phi(T_1) = 3H(T_1)\,.
\end{equation} 
If $T$ is below $\Lambda_h$ during reheating, oscillations start before reheating ends.
Then, the inflaton relic density from oscillation is roughly estimated by  
\begin{equation}
\label{eq:inflaton-relic}
\Omega_\phi h^2 \sim 
0.24\,\theta^2_c
\left( \frac{T_1}{\Lambda_h}\right)^n
\left(\frac{f}{10^{11}{\rm GeV}} \right)^2 
\left( \frac{T_{\rm reh}}{10^5{\rm GeV}} \right)\,,
\end{equation}
with $T_1<\Lambda_h$,  and $n=11N/6-2$ for confining SU$(N)$.
Here $T_{\rm reh}$ is the reheating temperature at which the Universe becomes completely 
radiation-dominated,  and we have used that the scale factor scales
as $H^{-2/3}$ during a matter-dominated era.
If oscillations start after reheating, the relic density can be read off from \eqref{eq:inflaton-relic} 
by replacing $T_{\rm reh}$ with $T_1$. 
$\phi$ can thus account for the observed dark matter in a wide range of $f$ 
depending on $T_{\rm reh}$. 
It is worth noting that $\alpha=0$ leads to accidental $Z_2$ even when 
\eqref{eq:DM} is not the case~\cite{Im:2019iwd}.

The inflation sector includes another candidate for dark matter associated with
spontaneously broken U$(1)_\chi$.   
An interesting possibility arising due to a wide allowed range of $H_{\rm inf}$ 
is to identify U$(1)_\chi$ with the PQ symmetry so that $\arg{\chi}$ 
becomes the QCD axion explaining the absence of CP violation in QCD.  
This implies that the waterfall and PQ phase transitions are identical. 
Then,   as corresponds to $\chi_0$, the axion decay constant is cosmologically determined by 
\begin{equation}
f_a 
\approx 
\frac{3.8\times 10^{11}\,{\rm GeV}}{ \lambda^{1/4} }
\left( \frac{H_{\rm inf}}{10^4{\rm GeV}} \right)^{1/2}
\, ,
\end{equation}
which should be above about $10^9$~GeV to avoid astrophysical bounds. 
The axion anomalous coupling to gluons,   which is required to solve the strong CP problem,  
is generated by adding U$(1)_\chi$-charged heavy quarks 
or extra Higgs doublets~\cite{Ringwald:2012hr,Kawasaki:2013ae}.
We also note that the domain-wall number should be equal to one since otherwise domain walls formed 
during the QCD phase transition overclose the Universe. 
In such a case,  axions are produced from coherent oscillations and,  more efficiently,  
from unstable domain-walls bounded by an axion string.   
The relic density is estimated by~\cite{Hiramatsu:2012sc} 
\begin{equation}
\Omega_a h^2 
\approx
0.54 \times 
\left( \frac{\Lambda_{\rm QCD}}{400{\rm MeV}} \right)
\left( \frac{f_a}{10^{11}{\rm GeV}} \right)^{1.19}
\, .
\end{equation}
Therefore, the observed dark-matter density indicates 
\begin{equation}
H_{\rm inf} \lesssim \sqrt\lambda\times 10^4\,{\rm GeV}\,.
\end{equation} 
It would be also interesting to consider other cases where U$(1)_\chi$ is identified, 
for instance, with U$(1)_L$ associated with the seesaw mechanism 
or local U$(1)_{B-L}$ to extend SM.

\section{CONCLUSIONS}
\label{sec:conclusion}

We have proposed a cosmological scenario that improves significantly the naturalness
and viability of axion-driven inflation.  
During inflation, the waterfall field remains at the origin by a potential barrier,   
which disappears when the inflaton reaches at a critical point--then inflation ends almost instantaneously.
The inflaton interaction responsible for such a barrier can naturally arise if the shift symmetry is 
broken nonperturbatively by hidden QCD with quarks coupled to the waterfall field. 
Interestingly, the Planck results indicate 
the possibility of probing our scenario by experimental searches for axionlike particles.
It is also remarkable that the inflaton can be stable enough to constitute dark matter 
if all the hidden quarks get heavier than the confining scale at the true vacuum.  
Further, for the case of a complex waterfall field, its phase component can play the role of 
the QCD axion, contributing to dark matter for $H_{\rm inf}$ below about $10^4$~GeV.

\subsection*{Acknowledgements}

This work is supported in part by the National Research Foundation of Korea Grants 
No.  2018R1C1B6006061,  No. 2021R1A4A5031460 (K.S.J.) and No.  2019R1A2C2085023 (J.G.).
We also acknowledge the Korea-Japan Basic Scientific Cooperation Program supported 
by the National Research Foundation of Korea and the Japan Society for the Promotion of Science 
(2020K2A9A2A08000097). 
J.G. ~is further supported in part by the Ewha Womans University Research Grant of 2020 (1-2020-1630-001-1). 
J.G.~is grateful to the Asia Pacific Center for Theoretical Physics for hospitality 
while this work was under progress.

\end{document}